\documentclass[astrosymb, twocolumn]{aastex701}
\usepackage{color, soul, ulem}
\usepackage{comment}
\usepackage{physics}
\usepackage{siunitx}
\usepackage{standalone}
\usepackage{tikz}
\usetikzlibrary{decorations.pathmorphing, patterns, shapes.geometric, positioning}
\def\half{\frac{1}{2}}
\def\third{\frac{1}{3}}
\def\kB{k_\text{B}}
\def\kps{\kappa_\text{scat}}
\def\kpa{\kappa_\text{abs}}
\def\kpl{\kappa_\text{line}}
\def\kpt{\kappa_\text{tot}}
\def\kpcore{\kappa_\text{core}}
\def\kpsx{\kappa^\text{scat}}
\def\kpax{\kappa^\text{abs}_x}

\def\kptx{\kappa^\text{tot}_x}
\def\kpanu{\kappa^\text{abs}_\nu}
\def\kptnu{\kappa^\text{tot}_\nu}
\def\Halpha{\mathrm{H}\alpha}
\def\Hminus{\mathrm{H}^-}
\newcommand{\parfrac}[2]{\left(\frac{#1}{#2}\right)}

\begin{document}

\title{Emission line formation in scattering dominated media: implications for LRDs}

\author[orcid=0000-0003-4105-3443]{Elisha Modelevsky}
\affiliation{Racah Institute of Physics, The Hebrew University, 91904, Jerusalem, Israel}
\email[show]{elisha.modelevsky@mail.huji.ac.il}

\author[orcid=0009-0001-0346-6434]{Omri Nitzan}
\affiliation{Racah Institute of Physics, The Hebrew University, 91904, Jerusalem, Israel}
\email{omri.nitzan@mail.huji.ac.il}

\author[orcid=0000-0002-1084-3656]{Re'em Sari} 
\affiliation{Racah Institute of Physics, The Hebrew University, 91904, Jerusalem, Israel}
\email{sari@phys.huji.ac.il}

\author[orcid=0000-0001-9185-5044]{Eliot Quataert} 
\affiliation{Department of Astrophysical Sciences, Princeton University, 4 Ivy Lane, Princeton, NJ 08540, USA}
\email{quataert@princeton.edu}

\begin{abstract}

Recent JWST observations of ``Little Red Dots'' (LRDs) reveal broad and prominent Balmer emission lines.
We present a theoretical framework for intrinsic emission line formation and broadening within static, optically thick, scattering-dominated gas envelopes with thermal populations. 
Using random-walk and diffusion approximations, we derive analytical line profiles for lines forming intrinsically within the scattering medium.
We demonstrate that a geometrically thin planar photosphere produces a shallow line profile characterized by a logarithmic plateau and a $v^{-1}$ wing.
A radially extended photosphere yields a broken power-law spectrum transitioning from $v^{-\alpha}$ to $v^{-(\alpha+1)}$, with \(0 < \alpha < 1\).
This is in contrast to a scattering medium external to the line-forming region, which produces an exponential line profile.
We show that this broken power-law model can fit the $\mathrm{H}\alpha$ line profiles observed in LRDs.
Higher quality spectra may be able to distinguish between intrinsic and extrinsic models for the line broadening in LRDs.
In our LTE models, the high contrast between the $\mathrm{H}\alpha$ and continuum flux cannot be explained.
Quantitative comparison to LRD spectra requires expanding our models to non-LTE situations.

\end{abstract}


\section{Introduction} 

JWST discovered a population of compact red objects called ``Little Red Dots'' (LRDs), mostly at high redshift \(z \sim 5\) \citep{Furtak2023,Matthee2024}.
LRDs' defining features are their compactness, a V-shaped spectrum with a break at the Balmer limit \citep{Setton2025}, and broad and prominent Balmer series emission lines, with \(\Halpha\) often accounting for \(\sim\)10\% of the luminosity \citep{deGraaff2025census}.
\added{The continuum emission of LRDs roughly fits blackbodies with temperatures \(T=3000-6000\,\text{K}\) \cite{deGraaff2026modBB}, and their luminosities are in the range \(10^{43}-10^{45}\,{\rm\frac{erg}{s}}\) \cite{Greene2026}.
The corresponding luminosity radius is \( R_\text{lum} = 10^{16}\,\text{cm} \parfrac{L}{10^{44}\,{\rm\frac{erg}{s}}}^\half \parfrac{T}{6000\,\text{K}}^{-2} \).
}

\added{
The Balmer line shapes of LRDs have been actively discussed as key diagnostics.
Recent studies reveal that the broad Balmer components frequently exhibit symmetric wings well-fit by either exponentials \citep{Rusakov2026, Brazzini2026} or Lorentzians \citep{Scholtz2026}.
}

Most current theories about their nature include an accreting massive black hole surrounded by dense gas \citep{InayoshiHo2025}, which must be Compton-thick enough to obscure X-rays \citep{Sacchi2025, Sneppen2026_quenching}.
Some models also use the dense gas envelope to explain line broadening through Thomson scattering \citep{Rusakov2026,Chang2026}\added{, while others invoke Keplerian kinematic broadening similarly to standard AGN \citep{Brazzini2025, Juodzbalis2026_directmass, Scholtz2026}.
The two interpretations are currently debated.}

\begin{figure*}
    \centering
    \resizebox{\linewidth}{!}{
        \begin{tikzpicture}[
    photon/.style={->, decorate, decoration={snake, amplitude=0.8pt, segment length=5pt}, shorten >=1pt, shorten <=1pt},
    region/.style={draw, circle, minimum size=3cm, align=center, font=\small\sffamily, thick},
    scatterer/.style={draw, ellipse, minimum width=4.5cm, minimum height=1cm, align=center, font=\small\sffamily, thick},
    opaque/.style={draw, rectangle, minimum width=1.2cm, minimum height=4.2cm, pattern=north east lines, thick},
    panel frame/.style={draw, thick},
    title/.style={font=\Large\sffamily, anchor=south}
]

\tikzset{
  observer/.pic={
    \draw[thick] (0,0) -- (0.6, 0.32);
    \draw[thick] (0,0) -- (0.6, -0.32);
    \draw[thick] (0.4, 0.24) arc (31:-31:0.46); 
  }
}

\begin{scope}[shift={(0,0)}]
    \draw[panel frame] (-3.75, -4) rectangle (3.75, 4);
    \node[title] at (0, 4.2) {\textbf{Backlit}};

    \pic[rotate=-45] at (-3.2, 3.5) {observer};

    \node[region] (agn1) at (0.5, -1.8) {Line forming\\region\\(AGN?)};
    \node[scatterer, rotate=30] (scat1) at (-0.5, 1) {Scatterer};

    \draw[photon] (agn1.150) -- (scat1.197);
    \draw[photon] (agn1.130) -- (scat1.210);
    \draw[photon] (agn1.110) -- (scat1.250);
    \draw[photon] (agn1.90)  -- (scat1.305);
    \draw[photon] (agn1.70)  -- (scat1.330);

    \draw[photon] (scat1.120) -- ++(-1.5, 1.5);
    \draw[photon] (scat1.150) -- ++(-1.8, 0.8);
    \draw[photon] (scat1.90)  -- ++(-0.5, 1.8);
    \draw[photon] (scat1.60)  -- ++(1.0, 1.5);
    \draw[photon] (scat1.20)  -- ++(1.5, 1.0);
    \draw[photon] (scat1.190) -- ++(-1.5, -1.0);
    \draw[photon] (scat1.200) -- ++(-0.5, -1.5);
\end{scope}

\begin{scope}[shift={(8.25,0)}]
    \draw[panel frame] (-4.5, -4) rectangle (4.5, 4);
    \node[title] at (0, 4.2) {\textbf{Reflected}};

    \pic[rotate=20] at (-4.2, 0) {observer};

    \node[region] (agn2) at (2.1, -1.8) {Line forming\\region\\(AGN?)};
    \node[scatterer] (scat2) at (2.1, 2.0) {Scatterer};
    
    \node[opaque] (opaq) at (-1.2, -1.7) {};
    \node[rotate=-90, font=\small\sffamily, fill=white, inner sep=2pt] at (-1.2, -1.7) {Opaque};

    \draw[photon] (agn2.120) -- (scat2.205);
    \draw[photon] (agn2.100) -- (scat2.235);
    \draw[photon] (agn2.80)  -- (scat2.305);
    \draw[photon] (agn2.60)  -- (scat2.335);

    \draw[photon] (scat2.028) -- ++(+0.5, +0.5);
    \draw[photon] (scat2.100) -- ++(+0.0, +0.5);
    \draw[photon] (scat2.160) -- ++(-0.5, +0.5);
    \draw[photon] (scat2.190) -- ++(-3.0, -1.0);
    \draw[photon] (scat2.198) -- ++(-1.2, -1.5);
\end{scope}

\begin{scope}[shift={(16.5,0)}]
    \draw[panel frame] (-3.75, -4) rectangle (3.75, 4);
    \node[title] at (0, 4.2) {\textbf{Intrinsic}};

    \pic[rotate=180+52] at (3, 3) {observer};

    \node[region, minimum size=3.5cm] (agn3) at (0, -1.5) {Line formation\\and scattering};

    \draw[photon] (agn3.40) -- ++(1.5, 1.5);
    \draw[photon] (agn3.60) -- ++(1.2, 1.8);
    \draw[photon] (agn3.80) -- ++(0.8, 2.0);
    \draw[photon] (agn3.20) -- ++(1.8, 1.0);
\end{scope}

\end{tikzpicture}
    }
    \caption{Schematic representation of backlit, reflected, and intrinsic line formation.
    \emph{Backlit}: the line is created in one region, and then passes through a scattering medium that broadens it.
    \emph{Reflected}: the line is created in one region, but is not directly visible to the observer, and is seen only after being scattered back by a scattering medium.
    \emph{Intrinsic}: the line is created and broadened in the same scattering medium.
    The intrinsic scenario is the simplest, requiring a single object to produce the line shape.
    }
    \label{fig:geom_schematic}
\end{figure*}
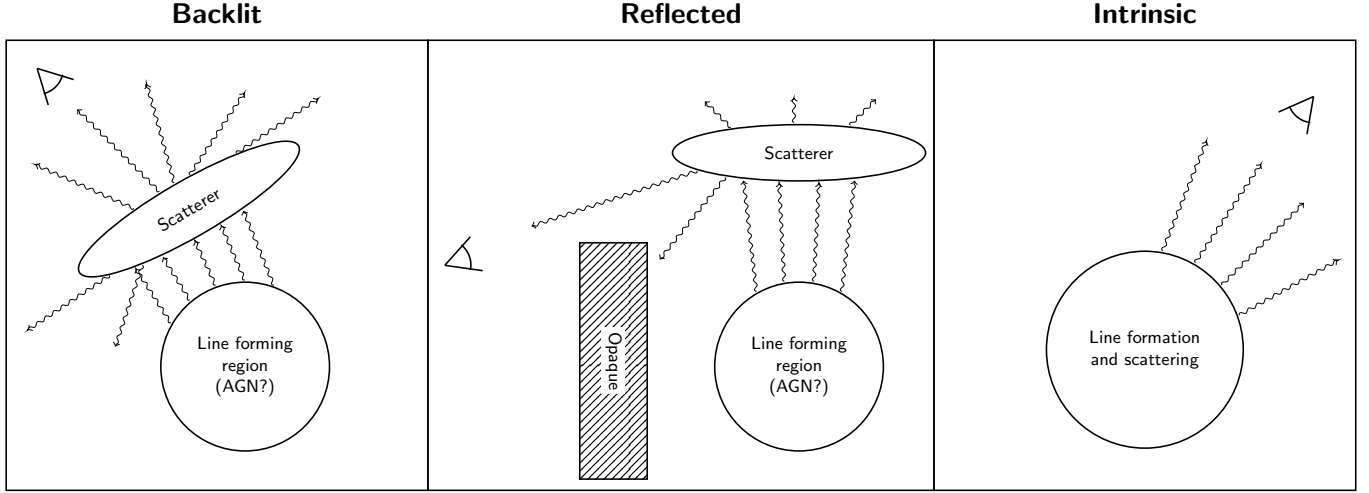

In Thomson-broadening models, the line is assumed to be created in one region and then pass through a scattering medium that broadens it, forming exponential line wings as described by \cite{Laor2006, Chugai2001, RothKasen2018} in the context of supernovae and tidal disruption events.
We refer to this as the ``backlit'' scenario, schematically illustrated in figure~\ref{fig:geom_schematic}.
A similar scenario, which we call ``reflected'', is when the line forming region is not directly visible to the observer (for example, due to an opaque torus);
the line is seen only after being scattered back by a surrounding medium.

A very different scenario is when the line is both created and broadened in the same scattering medium, which we call the ``intrinsic'' scenario.
This is relevant to LRD models where the gas is thick enough to obscure any light coming from the center of the object \citep{Kido2025, Begelman2026}.
In such models, the strong Balmer lines and the red continuum must be created in the gas envelope itself.
Previous works on intrinsic line formation in scattering-dominated media have assumed that line opacity is negligible \citep{Weymann1970,SunyaevTitarchuk1980,HuangChevalier2018}.
This is generally not the case for the density and temperature ranges estimated for LRD gas envelopes.
Since the nature of LRDs is still an open question, both extrinsic (backlit, reflected) and intrinsic scenarios are viable.
The line profile may allow us to distinguish between them.

In section~\ref{sec:extrinsic} we show that \added{backlit} scenarios always produce exponential line profiles.
In the rest of this paper, we focus on intrinsic line formation and broadening.
Section~\ref{sec:assumptions} details the model assumptions, and section~\ref{sec:mechanism} shows the line formation mechanism, calculating the line shape for a planar photosphere in LTE.
Section~\ref{sec:radially_extended} shows that a radially extended photosphere produces a broken power-law line shape, and section~\ref{sec:observations_comparison} demonstrates that such a line shape can fit LRD observations.

\section{Exponential line profile in backlit broadening}
\label{sec:extrinsic}

In this short section, we briefly explain why \added{backlit} line broadening scenarios always produce exponential line profiles, as observed in numerical simulations \citep{Chugai2001, RothKasen2018}.
Our treatment is similar to the simplified method described in \cite{Laor2006} for \(\tau \ll 1\), extending the argument for \(\tau \gg 1\) as well.

The main idea is that the probability that a photon experiences \(n\) scatterings scales as \(p_n \sim e^{-\alpha n}\) for large enough \(n\).
The value of $\alpha$ depends on the optical depth $\tau$.
In the \(\tau \ll 1\) limit (as explained by \cite{Laor2006}), given that \(n\) scatterings occurred, the probability of another one occurring is \(\sim\tau\);
so \(p_n \sim \tau^n = e^{-\log (\tau^{-1}) n}\), \(\alpha \sim -\log \tau \).

In the \(\tau \gg 1\) limit, photons perform a random walk in the scattering medium.
Moving a distance of \(\tau\) takes \(\sim \tau^2\) steps.
Every \(\tau^2\) steps, the probability of the photon remaining in the medium is reduced by a \(\mathcal{O}(1)\) factor, giving \(\alpha \sim \tau^{-2}\).
This expression continuously transitions to the \(\tau \ll 1\) expression.
These arguments provide the rough scaling of \(\alpha\) as a function of optical depth.
The proportionality factor depends on the specific geometry.

Assuming that a photon that scattered \(n\) times has a Gaussian spectral distribution around the line with variance \(\propto n\), the resulting flux spectrum will be
\begin{equation}
    \sum_{n=1}^\infty \frac{e^{-\frac{x^2}{2n} - \alpha n}}{\sqrt{2 \pi n}}
    \approx \int_0^\infty \dd{n} \frac{e^{-\frac{x^2}{2n} - \alpha n}}{\sqrt{2 \pi n}}
    = \frac{e^{-\sqrt{2\alpha} \, x}}{\sqrt{2\alpha}} ,
\end{equation}
where \(x = \frac{\nu - \nu_0}{\nu_0 \sqrt\frac{2 \kB T}{m_e c^2}} \) is the normalized frequency shift.
Appendix~\ref{sec:large_tau_backlit} analytically calculates the exact value of \(\alpha\) in the \(\tau \gg 1\) limit for a slab geometry scattering medium, yielding
\begin{equation}
    \alpha \approx \frac{\pi^2}{3 \tau^2} \implies 
    \text{line profile} \sim e^{-\sqrt\frac{2}{3} \frac{\pi x}{\tau}} .
\end{equation}
The exponential e-folding scale in terms of Doppler velocity is then
\begin{equation}
    \text{line} \sim e^{-\frac{v}{w}}
    \implies
    w \approx \tau \cdot215 {\rm \frac{km}{s}} \parfrac{T}{10^4\;{\rm K}}^\half .
\end{equation}
This is consistent with the simulation fit in \cite{Rusakov2026}.
Their simulated geometry was a thin spherical shell around a point source, which is similar to two slabs around the photon source (with a combined optical depth of \(2\tau\)).
Due to this difference in definitions, their value of \(\dv{w}{\tau}\) is twice our analytically derived value.

\section{Intrinsic model assumptions}
\label{sec:assumptions}

We assume:
\begin{itemize}
    \item Opacity hierarchy: continuum absorption \(\ll\) Thomson scattering \(\ll\) line absorption.
    \item Matter is in thermodynamic equilibrium with itself (i.e., atomic level populations and the ionization fraction are thermal).
    \item The medium is static.
    \item Recoil is negligible in the photon energy redistribution for Thomson scattering.
\end{itemize}

\begin{figure*}
    \centering
    \includegraphics[width=\textwidth]{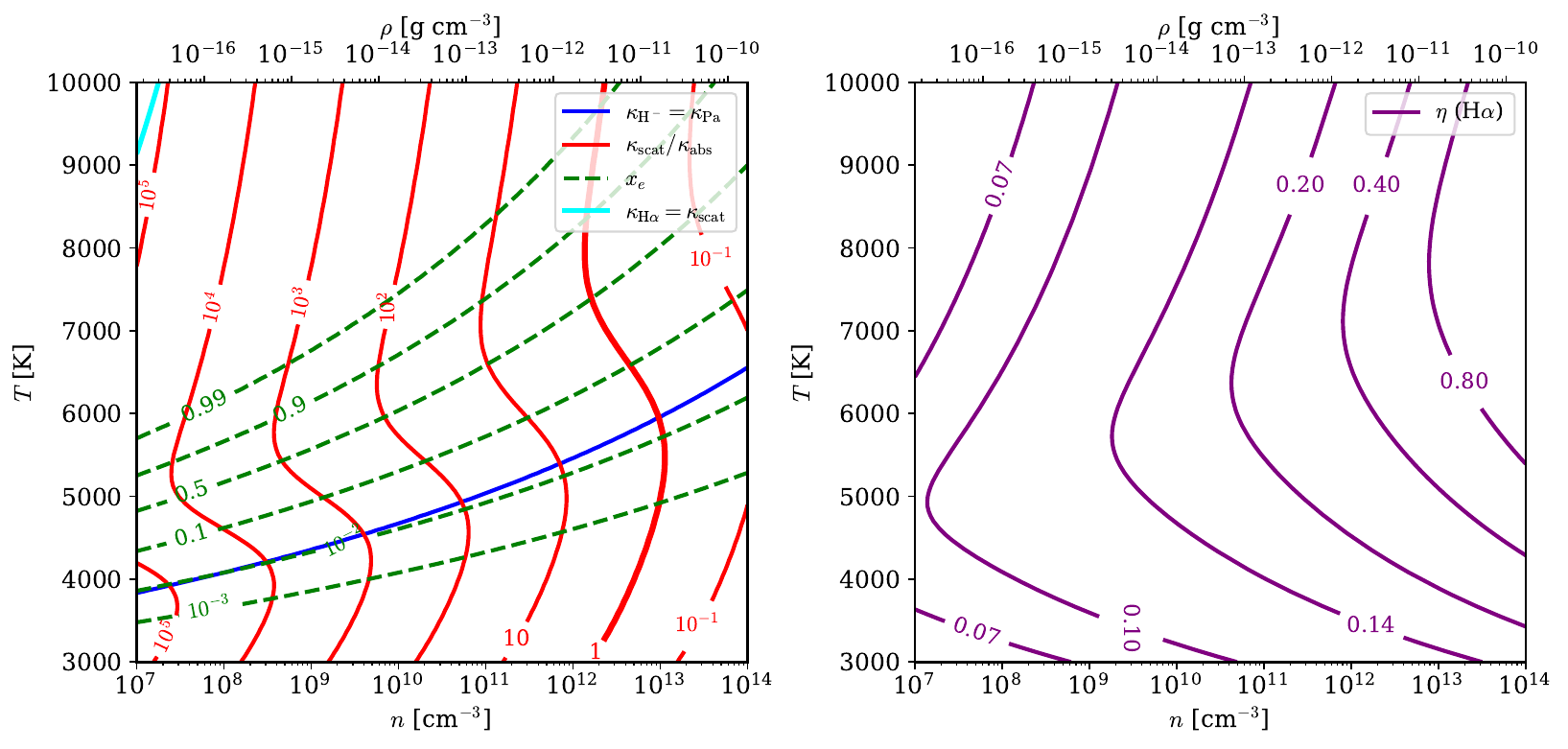}
    \caption{
    \emph{Left panel:}
    Atomic values as functions of density and temperature for pure hydrogen in LTE.
    In red, \(\frac{\kps}{\kpa}\) ratio for the continuum at 1.9 eV (near \(\Halpha\)).
    In blue, the line separating \(n=3\) photoionization and \(\Hminus\) photodetachment continuum opacity domination.
    In dashed green, the ionization fraction \(x_e\).
    In cyan, the line separating line opacity domination and Thomson scattering domination for \(\Halpha\) (almost all of parameter space is dominated by line opacity).
    \emph{Right panel:}
    The value of \(\eta = \int \frac{\kappa_\text{line}(x)}{\kappa_\text{line}(x) + \kps} \dd{x}\) for the \(\Halpha\) line, as a function of density and temperature for pure hydrogen in LTE.
    The atomic line data was taken from NIST \citep{NIST_ASD}.
    }
    \label{fig:hydrogen_beta_eta}
\end{figure*}

Figure~\ref{fig:hydrogen_beta_eta} shows that for a pure hydrogen gas, Thomson scattering is dominant over continuum absorption for low enough densities \(n \lesssim 10^{13}~\unit{cm^{-3}}\), and that the \(\Halpha\) opacity is dominant over Thomson scattering for a wide range of densities and temperatures.

As shown in appendix~\ref{sec:atomic_rates}, the photoionization rate generally dominates over radiative decay and collisional de-excitation in conditions where the opacity is dominated by Thomson scattering.
This means that energy levels are not necessarily thermally populated;
if the \(n\)-level photo-ionizing radiation is weak, then with LTE populations the recombination rate to level \(n\) exceeds the photoionization rate from that level.
This imbalance over-populates the \(n\) level, enhancing the source of \(n \to\) lower levels line photons, and possibly elevating the contrast between the line and continuum radiation.
However, for the sake of a simple model, we neglect this effect and leave its detailed study to a future work.

Neglecting the bulk motion of the medium means that this model does not capture the line shape in the range of \(\pm\) the bulk velocity around the line center.
P Cygni-like absorption features appear in 30-50\% of LRDs \citep{JuodzbalisAbsorption2026,Rusakov2026}, with velocities of a few hundred \({\rm \frac{km}{s}}\).
This is \(\lesssim \sqrt\frac{2\kB T}{m_e} = 400 {\rm \frac{km}{s}} \parfrac{T}{5000\text{ K}}^\half\) --- a single electron scattering rms shift.
The line shape at higher offsets from the center should not be affected by bulk motion.

Neglecting the recoil effect is justified by the following order-of-magnitude argument.
Thomson scattering has an average photon energy shift of \(\frac{\Delta E}{E} \sim \frac{E}{m_e c^2} \sim 10^{-6}\) set by the recoil effect, and a standard deviation of \(\sqrt\frac{2 \kB T}{m_e c^2} \sim 10^{-3}\) set by the thermal Doppler effect.
The number of scatterings \(N\) required for the recoil offset \(\sim 10^{-6} N\) to be comparable to the Doppler width \(\sim 10^{-3} \sqrt{N}\) is \(\sim 10^6\).
The number of scatterings scales roughly as the scattering optical depth squared \(\tau^2\);
so as long as \(\tau \ll 10^3\), recoil effects may be safely neglected.

\section{Emission line formation mechanism}
\label{sec:mechanism}

We define:
\begin{itemize}
    \item \(\tau =\) the scattering optical depth, measured from the vacuum surface.
    \item \(x = \frac{\nu - \nu_0}{\nu_0 \sqrt\frac{2 \kB T}{m_e c^2}}\) is the frequency offset from the line center, in units where a single scattering has \(\mathrm{Var}(\Delta x)=1\).
    \item \(\beta = \frac{\kpa}{\kpt}\) is the (small) ratio between continuum absorption opacity and total (absorption + scattering) opacity.
    \item \(\eta = \int \frac{\kpl(x)}{\kpt(x)} \dd{x}\) is a measure of the (small) width of frequencies where the line opacity is dominant over the scattering opacity.
\end{itemize}

A scattering-dominated (\(\beta \ll 1\)) atmosphere with temperature \(T\) will have a continuum emission flux reduced by a factor of \(\frac{4}{\sqrt3}\beta^\half\) compared to a blackbody\footnote{Note that this changes the luminosity radius estimate to \(R_\text{lum} = 2\cdot10^{16}\,\text{cm} \parfrac{\beta}{10^{-2}}^{-\frac{1}{4}} \parfrac{L}{10^{44}\,{\rm\frac{erg}{s}}}^\half \parfrac{T}{6000\,\text{K}}^{-2} \).}.
An atomic line whose line-center opacity is higher than the scattering opacity will have the peak emission flux of a true blackbody, thus forming an emission line.

As figure~\ref{fig:hydrogen_beta_eta} shows, in cases of interest we have \(\eta \ll 1\) --- the line-dominant spectral region is small in terms of the scattering width (electron thermal velocity).
In this limit, Thomson scattering is responsible for setting the line profile.
A photon that is emitted from the line has a probability \(\frac{\kps}{\kpt}\) of scattering out of the line.
Thus, the energy flux of photons that escape the line in a section with line-absorption optical depth \(\dd{\tau_\text{line}}\) is
\begin{equation}
\begin{split}
    \frac{\text{\small{scattered photon energy}}}{\text{\small{time}}\times\text{\small{area}}\times\text{\small{solid angle}}}
    &= \int \dd{\nu} B_\nu(T) \frac{\kps}{\kpt} \dd{\tau_\text{line}} \\
    &= \int \dd{\nu} B_\nu(T) \frac{\kpl}{\kpt} \dd{\tau}
\end{split}
\end{equation}
Assuming \(B_\nu(T)\) changes slowly over the line-dominant spectral region, we define \(B = B_{\nu_0}(T)\) and get
\begin{equation}
    \frac{\text{\small{scattered photon energy}}}{\text{\small{time}}\times\text{\small{area}}\times\text{\small{solid angle}}}
    = \eta B \sqrt\frac{2 \kB T}{m_ec^2} \dd{\tau} .
\end{equation}

\subsection{Simplest model: semi-infinite uniform atmosphere}
\label{subsec:flat_isothermal}

We start with the simplest configuration to illustrate the line formation and broadening mechanism.
The \(z>0\) half-space is filled with a gas that has uniform density \(\rho\) and temperature \(T\).

For simplicity, we consider a discrete scattering problem:
photons random-walking on a one-dimensional spatial grid.
In terms of the scattering optical depth coordinate \(\tau\), the variance of a single step is \(\mathrm{Var}(\Delta\tau)=\frac{2}{3}\) --- so the grid point \(k\) corresponds to an optical depth \(\tau = \sqrt\frac{2}{3} k\).

The energy flux of the scattered photons per unit frequency in \(x\) units is
\begin{equation}
\label{eq:simple_scattered_initial_sum}
    F_\text{scat}(x) = \sum_{k=1}^\infty \sqrt\frac{2}{3} \eta B \sum_{n=k}^\infty 2^{-n} \frac{k}{n} \binom{n}{\frac{n+k}{2}} G(n,x),
\end{equation}
where the \(k\)-sum is over the starting point of a photon, and the \(n\)-sum is over the number of steps (scatterings) the photon experiences before reaching the surface.
\(2^{-n} \frac{k}{n} \binom{n}{\frac{n+k}{2}}\) is the probability that a random walking photon that started at \(k\) reaches \(0\) for the first time in exactly \(n\) steps.
\(G(n,x)\) is the spectral distribution of a photon that has undergone \(n\) scatterings, also taking into account that every scattering that passes through \(x=0\) has a chance \(\eta\) of destroying the photon.

\begin{figure*}
    \centering
    \includegraphics[width=\linewidth]{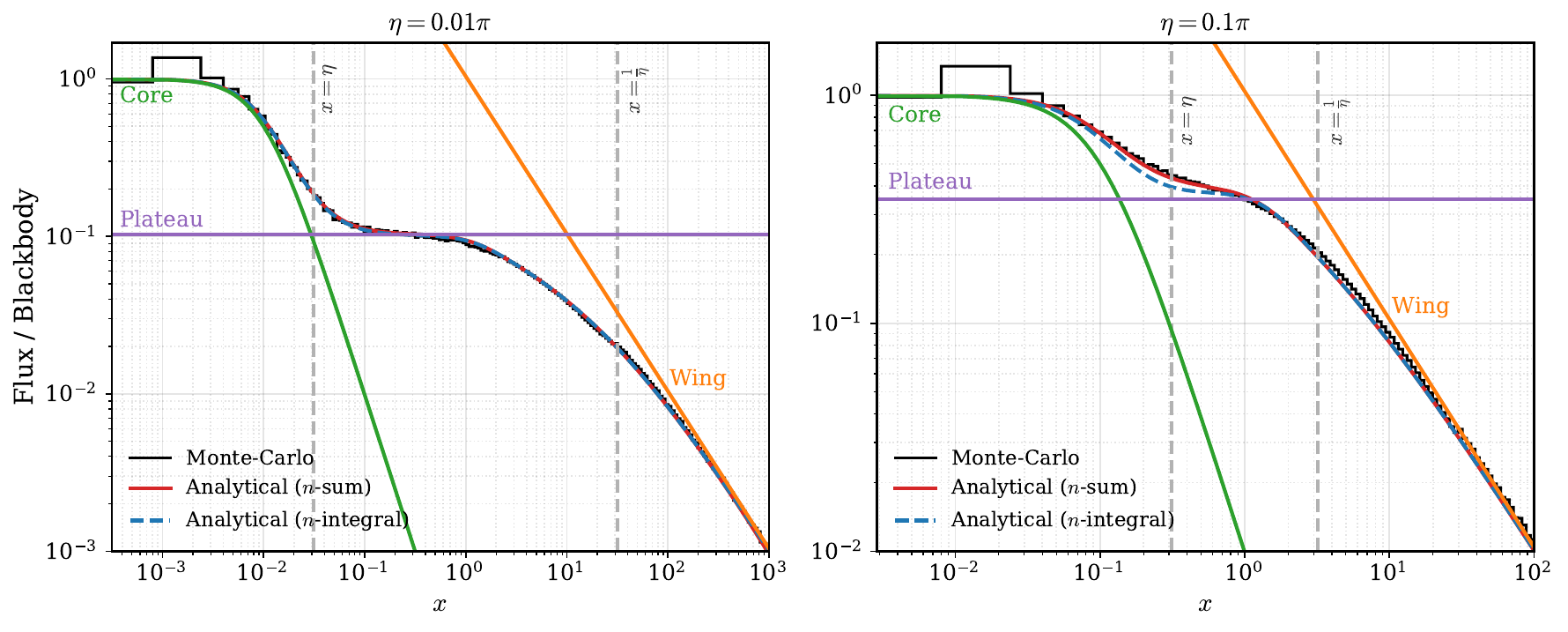}
    \caption{
    The line profile in planar geometry.
    A comparison of Monte Carlo simulations with the analytical line model (both the ``sum'' version from equation~\eqref{eq:simple_scattered_sum} and the approximated ``integral'' version from equation~\eqref{eq:simple_scattered_exp_integral}).
    The absorption opacity function is a Lorentzian tail \(\frac{\kpax}{\kpsx} = \parfrac{\eta/\pi}{x}^2\).
    Different components of the spectrum are also plotted;
    in green, the core (the non-scattered component, equation~\eqref{eq:non_scattered_component});
    in purple, the plateau (the \(x\to0\) limit of the scattered component, equation~\eqref{eq:simple_scattered_exp_integral});
    in orange, the wing (the \(x\to\infty\) limit of the scattered component, equation~\eqref{eq:simple_scattered_exp_integral}).
    Left: \(\eta = 0.01\pi\), right: \(\eta=0.1\pi\).
    }
    \label{fig:monte_carlo_vs_analytical_simple}
\end{figure*}

To simplify the expression, we use the combinatorial identity \(\sum_{k=1}^n \frac{k}{n} \binom{n}{\frac{n+k}{2}} = \binom{n-1}{\lfloor \frac{n-1}{2} \rfloor}\), and get
\begin{equation}
\label{eq:simple_scattered_sum}
\begin{split}
    F_\text{scat}(x) &= \sqrt\frac{2}{3} \eta B \sum_{n=1}^\infty G(n,x) 2^{-n} \binom{n-1}{\lfloor \frac{n-1}{2} \rfloor} \\
    &\approx \sqrt\frac{2}{3} \eta B \sum_{n=1}^\infty \frac{G(n,x)}{\sqrt{2 \pi n}} .
\end{split}
\end{equation}
The approximation is for \(n \gg 1\), but is never off by more than 20\% even for small \(n\).
To simplify this further, note that \(G(n,x)\) is the solution to a diffusion problem with a probability \(\eta\) sink at \(x=0\),
\begin{equation}
    \pdv{G}{n} = \half \pdv[2]{G}{x} - \eta \delta(x) G ,
\end{equation}
with the initial condition \(G(0,x) = \delta(x)\).
This equation is easy to solve in Laplace space
\begin{equation}
    \tilde{G}(s,x) = \int_0^\infty G(n,x) e^{-sn} \dd{n} = \frac{e^{-\sqrt{2s}|x|}}{\eta + \sqrt{2s}} .
\end{equation}
Approximating the sum as an integral using the Euler-Maclaurin formula,
\begin{equation}
\begin{split}
    F_\text{scat}&(x) = \sqrt\frac{2}{3} \eta B \times \\
    \Bigg[ & \int_0^\infty \frac{G(n,x)}{\sqrt{2 \pi n}} \dd{n} + \frac{G(1,x)}{2\sqrt{2\pi}}
    - \int_0^1 \frac{G(n,x)}{\sqrt{2 \pi n}} \dd{n} \Bigg]
\end{split}
\end{equation}
In the second and third term, we can use \(G(n \le 1,x) \approx \frac{e^{-\frac{x^2}{2n}}}{\sqrt{2 \pi n}}\) since the photon has not had a chance to be re-absorbed in the line.
For the first term, we use \(n^{-1/2} = \frac{1}{\sqrt\pi} \int_0^\infty s^{-1/2} e^{-sn} \dd{s}\) to make this an integral over \(\tilde{G}\) instead.
\begin{equation}
\begin{split}
    F_\text{scat}&(x) = \sqrt\frac{2}{3} \frac{\eta B}{\pi} \times \\
    \Bigg[ &
        \int_0^\infty \frac{e^{-\sqrt{2s}|x|}}{\eta + \sqrt{2s}} \frac{\dd{s}}{\sqrt{2s}}
        + \frac{e^{-\frac{x^2}{2}}}{4}
        - \int_0^1 \frac{e^{-\frac{x^2}{2n}}}{2n} \dd{n}
    \Bigg] .
\end{split}
\end{equation}
Both of these integrals are solvable in terms of the exponential integral
\(E_1(z) = \int_z^\infty \frac{e^{-t}}{t} \dd{t}\).
\begin{equation}
\label{eq:simple_scattered_exp_integral}
\begin{split}
    F_\text{scat}(x) &= \sqrt\frac{2}{3} \frac{\eta B}{\pi} \left[
        e^{\eta |x|} E_1(\eta |x|) + \frac{e^{-\frac{x^2}{2}}}{4} - \half E_1\left( \frac{x^2}{2} \right)
    \right] \\
    &= \sqrt\frac{2}{3} \frac{B}{\pi} \times
    \begin{cases}
        -\eta \frac{\gamma - \half + \log (2 \eta^2)}{2} & x \ll 1 \\
        -\eta \left( \gamma + \log (\eta |x|) \right) & \eta \ll \eta x \ll 1 \\
        \frac{1}{x} & 1 \ll \eta x
    \end{cases},
\end{split}
\end{equation}
where \(\gamma\approx0.577\) is the Euler-Mascheroni constant.
Finally, to get the actual line shape, the non-scattered photons emitted from the line core must also be taken into account.
Given that a photon with frequency \(x\) was not absorbed, the probability of it also not being scattered is \(\frac{\kpax}{\kptx}\).
Therefore, the non-scattered photon flux is a blackbody multiplied by this probability
\begin{equation}
\label{eq:non_scattered_component}
F_\text{non-scat} = \frac{\kpax}{\kptx} \frac{B}{4}.
\end{equation}
Similarly, the scattered component should be multiplied by the complementary probability \(\frac{\kpsx}{\kptx}\).
In total,
\begin{equation}
    F(x) = \frac{\kpax}{\kptx} \frac{B}{4} + \frac{\kpsx}{\kptx} F_\text{scat}(x)
\end{equation}
A comparison between this analytical model and a Monte Carlo simulation (see appendix~\ref{sec:monte_carlo} for details) is shown in figure~\ref{fig:monte_carlo_vs_analytical_simple}.

\subsection{Continuum absorption opacity}

Continuum absorption around the line has two main effects on the emitted spectrum:
\begin{enumerate}
    \item Continuum emission creates a ``baseline'' flux.
    \item Continuum absorption of line photons creates an exponential cutoff for the line profile.
\end{enumerate}
However, the exponential drop of the line is not significant;
as we will show, the exponential decay starts just where the line is comparable to the continuum baseline.

The production rate of continuum photons in LTE is
\begin{equation}
\begin{split}
    \frac{\small \text{emitted photon energy}}{\small \text{time}\times\text{area}\times\text{solid angle}}
    &= B \dd{\tau_\text{abs}} = B \frac{\kpa}{\kps} \dd{\tau} \\
    &= \frac{\beta}{1-\beta} B \dd{\tau}.
\end{split}
\end{equation}

Using the same formalism, we calculate the continuum emission
\begin{equation}
    F_\text{cont} = \sum_{k=1}^\infty \sqrt\frac{2}{3} \frac{\beta B}{1-\beta} \sum_{n=k}^\infty 2^{-n} \frac{k}{n} \binom{n}{\frac{n+k}{2}} (1 - \beta)^n
\end{equation}
where \((1-\beta)^n\) is the probability that a photon was not absorbed in the continuum after \(n\) scattering steps.
This double sum can be evaluated exactly,
\begin{equation}
    F_\text{cont} = \sqrt\frac{2}{3} \frac{\sqrt\beta}{\sqrt{\beta} + \sqrt{2-\beta}} B \approx \sqrt\frac{\beta}{3} B .
\end{equation}

To calculate the effect of continuum absorption on the line profile, we modify the sum in equation~\eqref{eq:simple_scattered_sum} by including a \((1-\beta)^n \approx e^{-\beta n}\) factor.
\begin{equation}
    F_\text{scat}(x) = \sqrt\frac{2}{3} \eta B \sum_{n=1}^\infty \frac{e^{-\beta n}}{\sqrt{2 \pi n}} G(n,x) .
\end{equation}
This line profile will differ from the \(\beta=0\) profile (equation~\eqref{eq:simple_scattered_exp_integral}) at frequencies \(x\) where the sum is dominated by \(n > \beta^{-1}\) terms.
Since most photons that reached \(x\) have experienced \(n \sim x^2\) scatterings, the exponential damping sets in at \(x \gtrsim \beta^{-1/2}\).
However, the \(\beta=0\) line is always \(\lesssim \frac{1}{x}\);
so at \(x \sim \beta^{-1/2}\) it will be similar to the continuum baseline \(\sim \beta^{1/2}\) or lower.
In other words, the exponential damping of the line begins when the line is barely above the continuum, making the actual exponential decay only observable after continuum subtraction.

\section{Radially extended photosphere}
\label{sec:radially_extended}

When the thickness of the photosphere of a spherical object is comparable to its radius, the surface area factor \(\sim r^2\) changes the relative importance of different layers.
Here we will examine the spectrum of a line formed in a spherically symmetric isothermal atmosphere where electron density is a power-law of radius \(n_e \propto r^{-\omega}\).
The scattering optical depth and the radius are related by
\begin{equation}
\label{eq:tau_r_powerlaw}
    \tau = \parfrac{r}{R_c}^{1-\omega},
\end{equation}
where \(R_c\) is the radius where \(\tau=1\).
For simplicity, we will assume that \(\eta\) is constant.
Figure~\ref{fig:hydrogen_beta_eta} shows that \(\eta\) indeed does not vary strongly wherever opacity is scattering-dominated.
However, if the density spans many orders of magnitude in the line emitting region, this assumption may be problematic.

\added{
We note that the planar geometry (considered in the previous section) can be thought of as the limit \(\omega\to\infty\), where the surface area factor \(r^2 \sim \tau^{-\frac{2}{\omega-1}} \sim \tau^0\) barely changes over arbitrary optical depth.
The non-uniform density in the \(\omega\to\infty\) limit does not matter for the analogy, since for the question of line formation and broadening only the source function \(\eta B\) per scattering optical depth matters.
}

\subsection{A schematic description of the line profile}
\label{subsec:schematic_line_nonflat}

We can get a rough picture of the line profile with a simple argument.
A photon with frequency shift \(x\) has most likely been scattered \(n \sim x^2\) times, and has been emitted from optical depth \(\tau\sim x\).
The emission rate per unit optical depth at \(\tau\) is \(4\pi r^2 \eta B\).
If \(\eta x \ll 1\), most photons do not get re-absorbed in the line.
If \(\eta x \gg 1\), the probability that a photon has reached \(x\) after \(n \sim x^2\) scatterings without passing through \(x=0\) more than \(\eta^{-1}\) times (and getting re-absorbed) is \(\sim \frac{1}{\eta x}\).
Therefore, the luminosity per unit frequency is roughly
\begin{equation}
\label{eq:rough_line_spectrum_changing_rho}
    L_\text{line}(x) \sim 4\pi R_c^2 B \times \begin{cases}
        \eta x^{-\frac{2}{\omega-1}} & \eta x \ll 1, x \gg 1 \\
        x^{-\frac{\omega+1}{\omega-1}} & \eta x \gg 1
    \end{cases}
\end{equation}
As detailed in appendix~\ref{sec:diffusion_solution}, the following interpolation formula works well when compared to Monte Carlo simulations:
\begin{equation}
\label{eq:interpolant_in_terms_of_omega}
    L_\text{scat} \approx 4 \pi R_c^2 B \frac{\eta C}{(1 + x)^{\frac{2}{\omega-1}} (\frac{2}{\omega-1} + \eta x)} ,
\end{equation}
where \(C = \frac{1}{\sqrt\pi} \parfrac{3}{2}^{\frac{1}{\omega-1}-\half} \frac{\Gamma\left(\frac{1}{\omega-1}+1\right)}{\Gamma\left(\frac{1}{\omega-1}+\half\right)}\).

Note that the source per unit \(\tau\) is \(4\pi r^2 \eta B \propto \tau^{-\frac{2}{\omega-1}}\), so if \(\omega<3\) most of the line photons are generated at \(\tau \ll 1\) (large radii), and the argument is invalid \added{since it relies on the diffusion approximation}.
Hence, equation~\eqref{eq:rough_line_spectrum_changing_rho} and equation~\eqref{eq:interpolant_in_terms_of_omega} are only valid for \(\omega > 3\);
for lower \(\omega\), a lower limit for \(\tau\) would have to be taken into account (such a limit may represent densities so low that the line is no longer dominant).

It is worth noting that a wind solution, which is a natural explanation for the prominence of P Cygni absorption lines in LRDs, does not necessarily have an electron power-law index \(\omega=2\).
One cause may be that the photosphere is in the accelerating region of the wind, where the density profile is steeper.
But even in the asymptotic \(\rho \sim r^{-2}\) region, the electron density profile is often steeper due to the ionization fraction decreasing outwards.
For example, in systems where ionization physics is similar to the Strömgren sphere \citep{Stromgren1939,McCullough2000}, the ionization fraction decreases with radius at the recombination front, causing a steeper electron density profile consistent with $\omega > 3$.
A recent example of an LRD model where this effect can be seen is the simulations in \cite{Sneppen2026_simulation}.
In systems where ionization is determined from Saha equilibrium, the ionization fraction also decreases outwards \citep{Strusberg2026, Yang2026}.
\added{The observed temperature of LRDs is \(\lesssim 6000\,\text{K}\), close to the ionization threshold at relevant densities (see figure~\ref{fig:hydrogen_beta_eta}).
Hence the ionization is expected to decrease with radius throughout the photosphere, and \(\omega>2\) is likely.}

\subsection{Continuum emission}

If the ratio of absorption to scattering opacities is constant, \(\beta=\text{const}\), then the emission rate per unit scattering depth is \(\sim 4\pi r^2 \beta B \propto \tau^{-\frac{2}{\omega-1}}\).
Once again,  we assume \(\omega > 3\) so that most of the photons are produced at \(\tau\gg1\).
If we define 
\begin{equation}
    \tau^* = \int_{r(\tau)}^\infty \sqrt{3 \kpa \kpt} \rho \dd{r} = \int_0^\tau \sqrt{3 \beta} \dd{\tau},
\end{equation}
then \(\tau^*\) is the effective absorption optical depth, taking into account the longer path a photon takes in a scattering-dominated atmosphere.
For \(\tau^* \gtrsim 1\), the emission is exponentially suppressed, so the total continuum emission is \(\sim \beta B R_c^2 \tau_m^{1 - \frac{2}{\omega - 1}}\), where \(\tau_m \sim \beta^{-1/2}\) is the scattering optical depth where \(\tau^*=1\).
Therefore, we get \(\sim B R_c^2 \beta^{\frac{\omega+1}{2\omega-2}} \) for the continuum baseline, which agrees with the power \(\beta^{1/2}\) for a semi-infinite space when \(\omega\to\infty\).

However, a constant \(\beta\) is not well-motivated physically.
Figure~\ref{fig:hydrogen_beta_eta} shows that taking \(\beta \propto \rho\) would be a better model.
Defining \(\beta_c\) as the value of \(\beta\) at \(\tau=1\), we get \(\beta = \beta_c \tau^{\frac{\omega}{\omega-1}}\).
Then \(\tau^* \sim \beta_c^{1/2} \tau^{\frac{3\omega-2}{2\omega-4}}\) and \(\tau_m \sim \beta_c^{-\frac{\omega-1}{3\omega-2}}\).
The emission per unit scattering optical depth is \(\sim \beta B r^2 \sim B R_c^2 \beta_c \tau^{\frac{\omega-2}{\omega-1}} \), so for \(\omega > \frac{3}{2}\) the total continuum emission is
\begin{equation}
\label{eq:rough_cont_spectrum_changing_rho}
    L_\text{cont} \sim 4\pi R_c^2 B \beta_c \tau_m^{1+\frac{\omega-2}{\omega-1}} \sim
    4\pi R_c^2 B \beta_c^{\frac{\omega+1}{3\omega-2}}
\end{equation}

\section{Comparison with LRD spectra}
\label{sec:observations_comparison}

We test whether this model can fit observed \(\Halpha\) lines in LRDs.
Formula~\eqref{eq:interpolant_in_terms_of_omega} for the scattered line profile in a power-law radially-extended photosphere can be written as
\begin{equation}
\label{eq:formula_to_fit}
    L(v) = \frac{A}{(1+\frac{v}{v_e})^\alpha (\alpha + \eta \frac{v}{v_e})} ,
\end{equation}
where \(v\) is the velocity offset from the line center, \(v_e=\sqrt\frac{2 \kB T}{m_e}\), \(\alpha = \frac{2}{\omega-1}\), and \(A\) is a normalization factor.
While in principle, all four parameters \(A,\alpha,\eta,v_e\) can be varied to fit data, the values of \(\eta\) and \(v_e\) are constrained by the red continuum blackbody fit and the plausible densities.

\begin{figure*}
    \includegraphics[width=\textwidth]{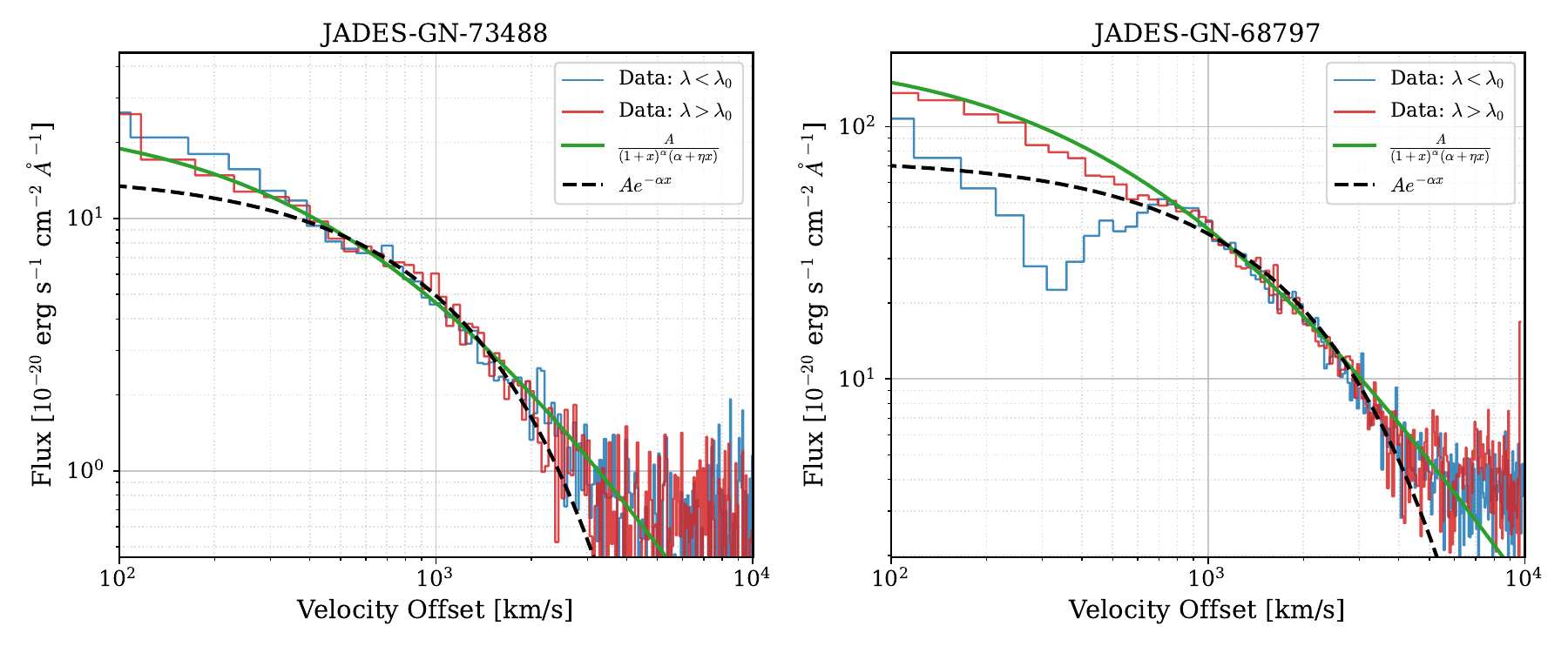}
    \caption{
    The \(\Halpha\) line in LRDs JADES-GN-73488 (left panel, \(z=4.13\)) and JADES-GN-68797 (right panel, \(z=5.04\)), flux as a function of absolute value velocity offset from the line center.
    The data is split into two profiles, one for each side of the line center (blue for blueshifted \(\lambda < \lambda_0\), red for redshifted \(\lambda > \lambda_0\)).
    Manual fits are shown, using equation~\eqref{eq:formula_to_fit} (green line) and an exponential \(A e^{-v/w}\) (dashed black line).
    Only parameters \(A\) and \(\alpha\) were fitted, while \(\eta=0.2\) and \(v_e=400\,\mathrm{km\,s^{-1}}\) were fixed to plausible values.
    The data was obtained from the DAWN JWST Archive (DJA) \citep{msaexp2023,deGraaff2025census,Heintz2024}.
    The chosen values of \(\alpha\) are \(\alpha=1.0\) for JADES-GN-73488 and \(\alpha=0.9\) for JADES-GN-68797.
    }
    \label{fig:jades}
\end{figure*}

Figure \ref{fig:jades} shows a manual ``by-eye'' fit of the \(\Halpha\) line in JADES-GN-68797 and JADES-GN-73488 (chosen because they have high resolution spectra) using equation~\eqref{eq:formula_to_fit}.
The values \(v_e=400~\mathrm{km\,s^{-1}}\) and \(\eta=0.2\) were fixed (corresponding to \(T=5000\,\mathrm{K}\) and \(n \sim 10^{11}\,\mathrm{cm^{-3}}\)), while \(A\) and \(\alpha\) were fitted.
Good fits are obtained with \(\alpha=0.9\) (for 68797) and \(\alpha=1.0\) (for 73488). 
Similarly good fits can be made for other LRDs with high signal-to-noise \(\Halpha\) lines.

It is worth noting that JADES-GN-68797 and JADES-GN-73488 are objects A and C (respectively) in \cite{Rusakov2026}, which showed that the \(\Halpha\) line is well-fit by an exponential profile.
We also show the exponential profile fit in figure~\ref{fig:jades}.
The good fit to a broken power-law is not necessarily surprising, considering \cite{Scholtz2026}, who showed that LRD data does not clearly favor exponential profiles over Lorentzian profiles.

A major caveat of the fit is that according to figure~\ref{fig:hydrogen_beta_eta}, the ratio between continuum absorption and scattering opacities at \(T=5000\,\mathrm{K},\;n \sim 10^{11}\,\mathrm{cm^{-3}}\) is \(\beta \sim 10^{-2}\).
The ratio between the line peak and the continuum is \(\lesssim \beta^{-\half} \sim 10\), which is inconsistent with the observed ratios of \(\sim 50\).
The thermalized-populations model thus cannot explain the observed line-to-continuum ratio.
However, this does not rule out the intrinsic line formation mechanism, since the line-to-continuum ratio is sensitive to the population of the excited levels --- which may be over-populated due to recombination.

The DAWN data products were processed using JWST Science Data Processing version 2025\_1 and the JWST Science Calibration Pipeline version 0.1.dev8704+g9b678d4.
Calibration reference files were selected using CRDS context jwst\_1298.pmap.

\section{Discussion}

We have shown how emission lines are formed intrinsically in Thomson scattering-dominated, optically thick objects, under the simplifying assumption of thermal populations.
When the physical thickness of the photosphere is small compared to the radius of the object, the scattered line profile can be divided into two parts:
\begin{itemize}
    \item The plateau \(x \ll \eta^{-1}\), where emission is roughly \(\sim \eta B\), and changes slowly (logarithmically) with \(x\) --- from repeated Thomson scatterings.
    \item The wing \(\eta^{-1} \ll x\), where emission is \(\sqrt\frac{2}{3} \frac{B}{\pi x}\) --- due to significant line re-absorption after more than \(\eta^{-1}\) Thomson scatterings.
\end{itemize}
\(x = \frac{\nu - \nu_0}{\nu_0 \sqrt\frac{2 \kB T}{m_e}}\) is the frequency shift normalized such that a single scatterings' shift has a variance of 1.

For a radially extended photosphere with \(n_e \sim r^{-\omega}\), the profile is modified to a broken power-law that steepens for large \(x\) due to line re-absorption:
\begin{itemize}
    \item The middle region \(x \ll \eta^{-1}\), where luminosity is roughly \( \sim \eta B R_c^2 (1+x)^{-\frac{2}{\omega-1}}\).
    \item The wing \(\eta^{-1} \ll x\), where luminosity is roughly \( \sim B R_c^2 x^{-\frac{2}{\omega-1} - 1} \).
\end{itemize}
Here, \(R_c\) is the radius where the scattering optical depth is unity, and the model is valid for \(\omega > 3\).
We note that even for an object with wind, it is plausible for the electron density profile to be steeper than \(r^{-2}\) due to the existence of a recombination front.

To fit the \(\Halpha\) line profiles in LRDs with this model, the photosphere must be radially extended;
the line profile of a geometrically thin photosphere is too shallow.
The line profile of our radially extended photospheres fits LRD observations well, but differs from the commonly assumed exponential profile (which is only appropriate for a line producing region external to the scatterer).
However, we find that the line-to-continuum contrast in our LTE  model is too low to explain observed LRD spectra, specifically the ratio of the Hydrogen Balmer line fluxes to the continuum.

A potentially important effect that was not explored in this work is the effect of non-LTE photoionization on the excited atom populations.
The more diluted the radiation is, the more over-populated the excited levels are, due to the recombination rate exceeding the photoionization rate.
Therefore, the low \(\tau\) layers may exhibit increased line production.
This could both mimic the profile steepening effect of a radially extended photosphere even for a geometrically thin photosphere, and increase the line-to-continuum contrast.

A full model for the spectrum of scattering-dominated photospheres will also need to include temperature gradients and outflow velocities.
In particular, the P Cygni-like absorption features that are found in many LRDs most likely require bulk motion in the photosphere.

\added{
Finally, we note how rotation may influence the line shape in the context of LRDs.
Taking \(L=10^{44}\,{\rm\frac{erg}{s}}\), \(T=6000\,\text{K}\) and \(\frac{\kpa}{\kps}=10^{-2}\), the luminosity radius is \(2\cdot10^{16}\,\text{cm}\).
Assuming Eddington luminosity, the mass is \(\sim 10^6\,{\rm M_\odot}\), giving a circular rotation speed of \(800\,{\rm \frac{km}{s}}\) at the luminosity radius.
For a pressure-supported structure, like a quasi-star, the rotation speed should be significantly slower, and it will not affect the broad line shape much.
For a kinematically-supported structure, the broad line features should be further smoothed by this \(\sim800\,{\rm\frac{km}{s}}\) speed scale, creating a somewhat broader line core, without a significant change to the line wings.
}


\begin{acknowledgments}
We thank Kenta Hotokezaka and Ari Laor for helpful discussions.
EM gratefully acknowledges support from the Milner Fellowship. 
EQ thanks Jenny Greene for getting him excited about LRDs. 
This research was partially supported by an NSF/BSF grant and a GIF grant.
This work is based in part on observations made with the NASA/ESA/CSA James Webb Space Telescope.
The primary archive of JWST data is the Mikulski Archive for Space Telescopes at the Space Telescope Science Institute, which is operated by the Association of Universities for Research in Astronomy, Inc., under NASA contract NAS5-03127 for JWST.
These observations are associated with program \#1181, and are accessible via \dataset[https://doi.org/10.17909/yh63-4396]{https://doi.org/10.17909/yh63-4396}.
The data products presented herein were retrieved from the Dawn JWST Archive (DJA).
DJA is an initiative of the Cosmic Dawn Center (DAWN), which is funded by the Danish National Research Foundation under grant DNRF140.
Support for US investigators in Program \# JWST-AR-10935.002-A was provided by NASA through a grant from the Space Telescope Science Institute, which is operated by the Association of Universities for Research in Astronomy, Inc., under NASA contract NAS5-03127.
\end{acknowledgments}

\appendix

\section{Calculation of atomic rates}
\label{sec:atomic_rates}

\subsection{Collisional de-excitation rate}

The de-excitation rate coefficient for the transition from the energy level \(n=u\) down to \(n=l\) is \citep{CloudyDatabase2015}
\begin{equation}
    q_{ul} = \num{8.629e-6}~\unit{\frac{cm^3}{s}} \parfrac{T}{1~\unit{K}}^{-1/2} \frac{\varUpsilon_{ul}}{g_u} ,
\end{equation}
where \(g_u = 2u^2\) is the degeneracy of the \(u\) level, and \(\varUpsilon_{ul}\) is the thermally averaged collision strength, which we approximate using the \cite{vanRegemorter1962} formula
\begin{equation}
    \varUpsilon_{ul} = \frac{8\pi}{\sqrt{3}} f_{ul} g_l \frac{I_H}{\Delta E_{ul}} \bar{g},
\end{equation}
where \(f_{ul}\) is the absorption oscillator strength of the transition line, \(I_H = 13.6~\unit{eV}\) is the ionization energy of hydrogen, \(\Delta E_{ul}\) is the transition energy, and \(\bar{g}\) is the effective Gaunt factor, which is given by
\begin{equation}
    \bar{g} = \max\left( 0.2, \frac{\sqrt{3}}{2\pi} e^{\frac{\Delta E_{ul}}{\kB T}} E_1\parfrac{\Delta E_{ul}}{\kB T} \right).
\end{equation}
\(E_1(z) = \int_z^\infty \frac{e^{-t}}{t} \dd{t}\) is the exponential integral function.
The total collisional de-excitation rate from the \(u\) level is then
\begin{equation}
    \text{thermal rate} = n_e \sum_{l=1}^{u-1} q_{ul} .
\end{equation}
The total rate is dominated by the \(u \to u-1\) transition, since up to the Gaunt factor \(\bar{g}\) which does not change very strongly,
\begin{equation}
    q_{ul} \propto \frac{f_{ul} g_l}{\Delta E_{ul}} \propto f_{ul} \frac{l^4}{u^2 - l^2}
\end{equation}
which is maximal for \(l=u-1\).

\subsection{Radiative decay rate}

A naive calculation of the decay rate of the \(n=u\) level to the \(n=l\) level would be the Einstein coefficient \(A_{ul}\).
However, when the photon is emitted by this decay, it may be re-absorbed by an \(l\)-level atom, re-creating a \(u\)-level atom.
Therefore, the actual decay rate would be \(A_{ul}\) divided by the expected number of emissions and re-absorptions, before another process ``destroys'' the line photon.

Often, such a process would be the physical escape of the photon from the medium \citep{RybickiLightmanCh1};
then the effective decay rate would be \(\frac{A}{1+\tau_\text{core}^2}\), where \(\tau_\text{core}\) is the optical depth of the medium for a line-core photon, and \(\tau_\text{core}^2\) is the expected number of resonant scatterings for the photon to traverse a distance of \(\tau_\text{core}\).

However, if the medium is optically thick to Thomson scattering, another ``photon-destruction'' process is dominant.
When a line photon undergoes Thomson scattering, its frequency shifts by \( \frac{\Delta\nu}{\nu} \sim \sqrt\frac{\kB T}{m_e c^2} \).
Due to the line Doppler width being set by the atomic thermal velocity, which is smaller than the electron thermal velocity by a factor of \(\sqrt\frac{m_p}{m_e} \approx 40\), often this shift is enough for the line absorption opacity to be lower than the Thomson scattering opacity.

The chance of a Thomson scatter occurring before re-absorption for a frequency \(\nu\) photon is \(\frac{\kpsx}{\kpsx + \kpanu} \).
The chance of a frequency \(\nu\) photon being created in the first place is proportional to \( \kpanu \).
Thus, the probability of Thomson scattering before re-absorption is
\begin{equation}
    P_\text{Thomson} = \frac{\int_{-\infty}^{+\infty} \frac{\kpanu}{\kptnu} \dd{\nu}}{\int_{-\infty}^{+\infty} \frac{\kpanu}{\kpsx} \dd{\nu}}
    \approx \sqrt\frac{m_p}{\pi m_e} \eta \frac{\kps}{\kpcore}
    = 24 \eta \frac{\kps}{\kpcore} ,
\end{equation}
if we assume that the thermal Doppler width is much greater than the natural line width, so for the integral in the denominator \(\kpanu \approx \kpcore \exp\left[-\frac{\frac{\nu-\nu_0}{\nu_0}}{2 \sqrt\frac{\kB T}{m_p c^2}}\right] \).
Hence, the total effective radiative decay rate is
\begin{equation}
    \text{decay rate} = \sum_{l=1}^{u-1} \frac{A_{ul}}{1 + \frac{\kpcore^{ul}}{24 \eta_{ul} \kps}}
\end{equation}
This rate is also dominated by the \(u \to u-1\) transition, since \(A_{ul}\) does not depend strongly on \(l\), whereas the line opacity \(\kpcore\) is smallest for \(l=u-1\) --- mainly due to the smaller population in higher energy levels.

\subsection{Photoionization rate}

The LTE photoionization rate from an atom at energy level \(n\) is \citep{Seaton1959,Nahar2021}
\begin{equation}
    \Gamma_n^\text{LTE} = \int_{\nu_n}^\infty \frac{8\pi \nu^2}{c^2} \frac{\sigma_n(\nu)}{e^\frac{h\nu}{\kB T} - 1} \dd{\nu} ~, \quad
    \sigma_n(\nu) = \frac{64\pi \alpha a_0^2}{3\sqrt3} g_\text{bf} n \parfrac{\nu}{\nu_n}^{-3} ~,
\end{equation}
where \(\nu_n\) is the ionization threshold frequency, \(g_\text{bf}\) is the Kramers-Gaunt factor (which we take to equal 0.9 for simplicity), \(\alpha\) is the fine structure constant and \(a_0\) is the Bohr radius.
To get a simpler formula, a useful approximation is
\begin{equation}
    \int_1^\infty \frac{\dd{t}}{t(e^{yt}-1)} \approx \frac{e^{-y}}{y} ,
\end{equation}
which is never off by more than 22\% --- so useful for order of magnitude calculations.
This gives
\begin{equation}
    \Gamma_n^\text{LTE} \approx \frac{512\pi^2}{3\sqrt3} \frac{\alpha a_0^2 \nu_1^2 \kB T}{hc^2} g_\text{bf} n^{-3} e^{-\frac{h \nu_1}{\kB T n^2}}
    = 2.7\cdot10^8~{\rm s^{-1}} \parfrac{T}{6000~{\rm K}} n^{-3} e^{-\frac{26.3}{n^2}  \parfrac{T}{6000~{\rm K}}^{-1}}
\end{equation}

\begin{figure*}
    \centering
    \includegraphics[width=\textwidth]{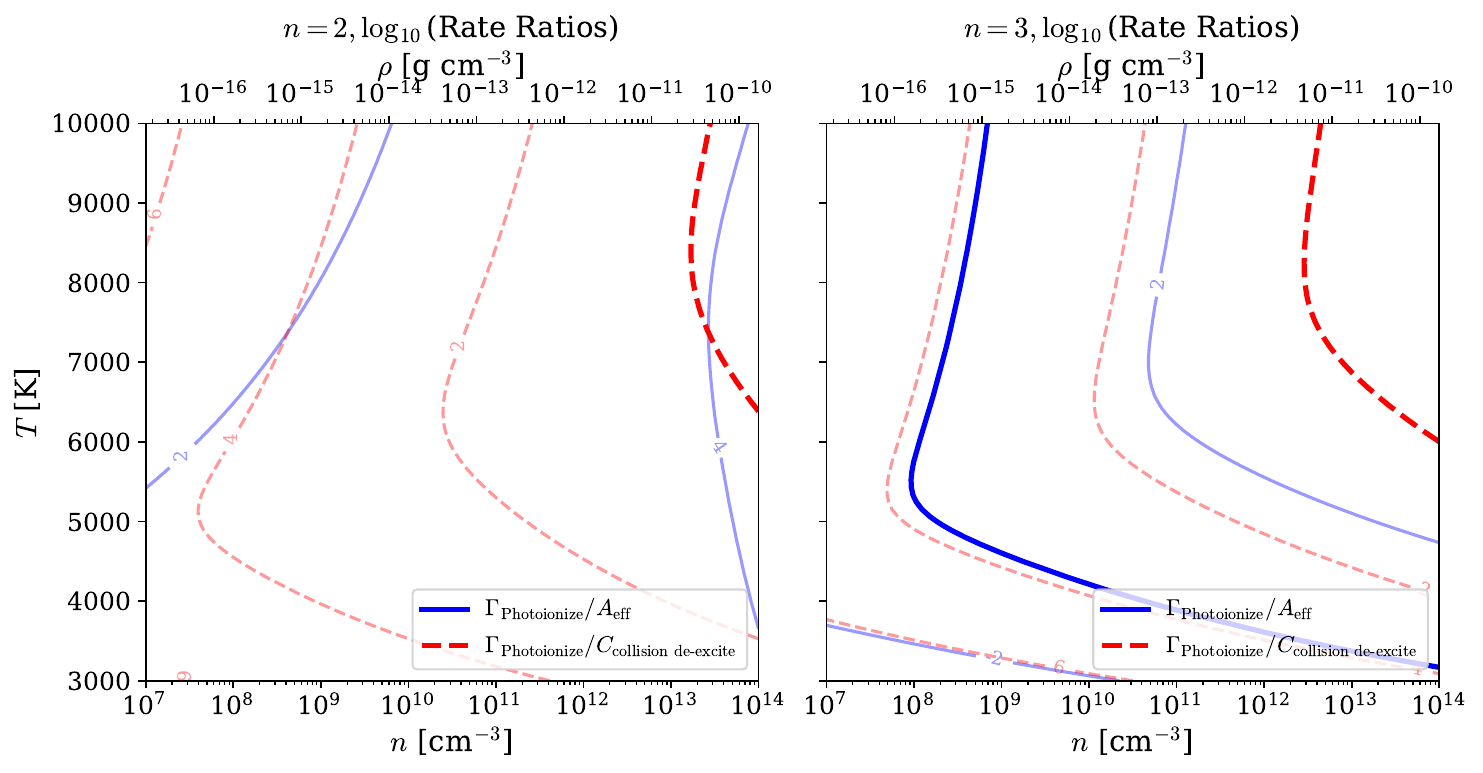}
    \caption{
    \(\log_{10}\) of the ratio between the LTE photoionization rate and the effective radiative decay rate (in blue), and the ratio between the LTE photoionization rate and the collisional de-excitation rate (in dashed red), for the \(n=2\) (left panel) and \(n=3\) (right panel) and \(n=4\) (green) energy levels.
    The ratios are plotted as contours in density and temperature.
    The bold lines are where the rates are equal.
    The assumptions are detailed in appendix~\ref{sec:atomic_rates}.
    The atomic line data was taken from NIST \citep{NIST_ASD}.
    }
    \label{fig:atomic_processes}
\end{figure*}

The ratio between the microscopic rates in LTE (photoionization, effective radiative decay and collisional de-excitation) is plotted in figure~\ref{fig:atomic_processes}.
Both for the \(n=2\) and \(n=3\) energy levels, the photoionization rate dominates for most of parameter space: \(3\cdot10^3\text{ K} < T < 10^4\text{ K}\), \(10^7~{\rm cm^{-3}} < n < 10^{14}~{\rm cm^{-3}}\).

\section{Backlit line broadening in the optically thick limit}
\label{sec:large_tau_backlit}

\cite{Laor2006} found an analytical solution for the scattered line shape in the backlit scenario (see figure~\ref{fig:geom_schematic}), in the optically thin scatterer limit.
We show here an analytical solution for an optically thick scatterer, in the specific case of a slab geometry.

Consider a monochromatic source of photons with flux \(F_\text{in} (x) = C \delta(x)\), where \(x = \frac{\nu - \nu_0}{\nu_0 \sqrt\frac{2 \kB T}{m_e c^2}}\) is the normalized frequency shift used throughout this work, hitting a slab of Thomson scattering optical depth \(\tau_0 \gg 1\) perpendicularly.
The radiative energy density \(J\) can be written as
\begin{equation}
\label{eq:backlit_decomposition}
    J(\tau,x) = \sum_{n=0}^\infty \psi_n(x) j_n(\tau) ,
\end{equation}
where \(\psi_n(x) = \frac{e^{-\frac{x^2}{2n}}}{\sqrt{2\pi n}}\) is the (approximate) spectrum after \(n\) scatterings and \(j_n(\tau)\) is the spatial distribution of \(n\)-scattered photons.
In the \(n \gg 1\) limit, the spatial distribution \(j_n(\tau)\) obeys the time-dependent diffusion equation
\begin{equation}
    \pdv{j}{n} = \third \pdv[2]{j}{\tau} .
\end{equation}
We solve this by expanding on eigenfunctions \(\pdv{j_k}{n} = -\alpha_k j_k\), where \(\alpha_k\) are eigenvalues to be determined.
Writing \(j_k(n,\tau) = e^{-\alpha_k n} u_k(\tau)\),
\begin{equation}
    \dv[2]{u_k}{\tau} = -3\alpha_k u_k \implies 
    u_k(\tau) = A_k \sin\left( \sqrt{3 \alpha_k} \tau + \phi_k \right) .
\end{equation}
The vacuum boundary conditions at \(\tau =0,\tau_0\) are \(\pdv{u}{\tau} = \pm \sqrt3 u\) (\(+\) at \(\tau=0\), \(-\) at \(\tau=\tau_0\)), yielding
\begin{equation}
    \tan \phi_k = \sqrt{\alpha_k} ~, \quad 
    \tan \left( \sqrt{3\alpha_k}\tau_0 + \phi_k \right) = -\sqrt{\alpha_k} ~.
\end{equation}
In the \(\tau_0 \gg 1\) limit, it takes many scatterings for photons to escape the slab --- so \(\alpha\) should be a small number for the leading-order eigenfunction.
Therefore, it is justified to approximate \(\phi_k \approx \sqrt{\alpha_k}\).
From this, we get
\begin{equation}
    \alpha_k = \frac{\pi^2}{3} \parfrac{k}{\tau_0 + \frac{2}{\sqrt3}}^2 .
\end{equation}
Plugging in the eigenfunction expansion into equation~\eqref{eq:backlit_decomposition} to get the outgoing flux,
\begin{equation}
\begin{split}
    F_\text{out} &= \frac{1}{\sqrt3} J(\tau=\tau_0) = \frac{1}{\sqrt3} \sum_{k=1}^\infty \sum_{n=0}^\infty \frac{A_k \sin \left( \sqrt{3\alpha_k} \tau_0 + \phi_k\right) }{\sqrt{2 \pi n}} \exp \left[ -\frac{x^2}{2n} - \alpha_k n \right] \\
    & \approx \sum_{k=1}^\infty \int_0^\infty  \dd{n} \frac{A_k (-1)^{k-1} \sqrt{\alpha_k}}{\sqrt{6 \pi n}} \exp \left[ -\frac{x^2}{2n} - \alpha_k n \right]
    = \frac{1}{\sqrt6} \sum_{k=1}^\infty A_k (-1)^{k-1} e^{-\sqrt{2\alpha_k} x} ~.
\end{split}
\end{equation}
To calculate \(A_1\), we need to expand the unscattered spatial distribution \(C e^{-\tau}\) in terms of the eigenfunctions.
The expansion is
\begin{equation}
    C e^{-\tau} = \sum_{k=1}^\infty A_k \sin\left( \sqrt{3\alpha_k} \tau + \phi_k \right) ~, \quad 
    A_k = C\cdot 2\pi\left(1+\frac{1}{\sqrt3}\right) \frac{k}{\left(\tau_0 + \frac{2}{\sqrt3}\right)^2} ~,
\end{equation}
valid in the limit \(k \ll \tau_0\).
From this, the outgoing flux is
\begin{equation}
    F_\text{out} = \frac{A_1}{\sqrt{6}} \sum_{k=1}^\infty (-1)^{k-1} k e^{-\sqrt{2\alpha_1} x k}
    = \frac{A_1}{\sqrt{6}} \frac{e^{-\sqrt{2\alpha_1}x}}{\left( 1 + e^{-\sqrt{2\alpha_1}x} \right)^2} .
\end{equation}
Denoting \(\tau_e = \tau_0 + \frac{2}{\sqrt3}\),
\begin{equation}
    F_\text{out} = C \frac{2\pi\left(1+\frac{1}{\sqrt3}\right)}{\sqrt6 \tau_e^2}
    \frac{e^{-\sqrt\frac{2}{3} \frac{\pi x}{\tau_e}}}{\left( 1 + e^{-\sqrt\frac{2}{3} \frac{\pi x}{\tau_e}} \right)^2} .
\end{equation}
The frequency-integrated flux is \(\sim \frac{C}{\tau_0}\) --- that is, a fraction \(\sim \tau_0^{-1}\) of incoming photons exit through the far side.
If we write the far \(x\) line shape as \( \propto e^{-\frac{v}{w}} \), where \(v = \sqrt\frac{2\kB T}{m_e} x\) is the velocity offset from the line center, then the exponential width is
\begin{equation}
    w = \sqrt\frac{2 \kB T}{m_e} \sqrt\frac{3}{2} \frac{\tau_0 + \frac{2}{\sqrt3}}{\pi}
    = (215 \tau_0 + 248) {\rm \frac{km}{s}} \parfrac{T}{10^4~{\rm K}}^\half
\end{equation}

Note that this value of \(\dv{w}{\tau}\) is a factor of 2 less than the fitted value in \cite{Rusakov2026}.
This is because the simulated geometry in \cite{Rusakov2026} was a thin spherical shell (\(\Delta R/R = 0.1\)), where reflected photons are never lost --- they reach the other side of the shell.
This is similar to a slab geometry where the optical depth is doubled, and the photons originate in the center of the slab instead of outside.

\section{A general diffusion model for the intrinsically formed line shape}
\label{sec:diffusion_solution}

In the diffusion approximation for both space and frequency, the radiative transfer equation in planar symmetry is
\begin{equation}
    \half \pdv[2]{J}{x} + \third \pdv[2]{J}{\tau} = \beta (J - B),
\end{equation}
where \(J\) is the radiative energy density, \(B\) is the blackbody energy density function, \(x = \frac{\nu - \nu_0}{\nu_0\sqrt\frac{2 \kB T}{m_e c^2}}\) is the frequency shift normalized for Thomson scattering, \(\tau\) is the total optical depth in the direction perpendicular to the symmetric plane, and \(\beta = \frac{\kpa}{\kpt}\).
\(B = B_\nu(T)\) may be a function of \(\tau\) for non-isothermal problems (we assume that the changes in \(x\) are too small to amount to a change in \(B\)), and \(\beta\) may be a function of both \(\tau\) and \(x\).

In spherical symmetry, the spatial Laplace operator gives
\begin{equation}
    \half \pdv[2]{J}{x} + \frac{1}{3r^2} \pdv{}{\tau} \left( r^2 \pdv{J}{\tau} \right) = \beta (J - B),
\end{equation}
From now on, we will restrict ourselves to power-law relations between \(\tau\) and \(r\)
\begin{equation}
    r = R_c \tau^{-p},
\end{equation}
like in equation~\eqref{eq:tau_r_powerlaw} with \(p = \frac{1}{\omega-1}\).
This gives
\begin{equation}
\label{eq:general_diffusion}
    \half \pdv[2]{J}{x} + \frac{\tau^{2p}}{3} \pdv{}{\tau} \left( \tau^{-2p} \pdv{J}{\tau} \right) = \beta (J - B) ,
\end{equation}
which can be used for the planar case as well by setting \(p=0\).
This diffusion equation can be further simplified by assuming that absorption opacity only exists in a narrow region around \(x=0\) (the line), so
\begin{equation}
    \beta(x) = \eta \delta(x), \quad 
    \eta = \int_{-\infty}^{+\infty} \beta(x) \dd{x}.
\end{equation}
In that case, for any \(x \ne 0\),
\begin{equation}
    \half \pdv[2]{J}{x} + \frac{\tau^{2p}}{3} \pdv{}{\tau} \left( \tau^{-2p} \pdv{J}{\tau} \right) = 0 .
\end{equation}
Integrating equation~\eqref{eq:general_diffusion} around \(x=0\) with the assumption of symmetry \(x \to -x\) gives a boundary condition at \(x = 0\):
\begin{equation}
\label{eq:J_BC_at_x0}
    \left. \pdv{J}{x} \right|_{x=0} = \eta ( J - B ) .
\end{equation}

To fully specify the spatial boundary value problem, we require boundary conditions at the deep interior ($\tau \to \infty$) and the outer surface ($\tau \to 0$).
In the deep interior \(J\) thermalizes, so \(J(\tau\to\infty,x)=B\).
For simplicity, we choose \(J(\tau=0,x)=0\) as the surface boundary condition --- a perfect sink.
While this is not as accurate as the standard \(J=\frac{1}{\sqrt3}\pdv{J}{\tau}\), the difference in the outgoing flux is only appreciable for optically thin photospheres.
Since we deal with optically thick objects (to Thomson scattering), this does not matter.

The total specific luminosity is obtained by integrating the flux \(\third \pdv{J}{\tau}\) over the spherical surface area and the outward solid angle:
\begin{equation}
\label{eq:luminosity_from_u}
    L(x) = \frac{4\pi}{3} \left[ r^2 \pdv{J}{\tau} \right]_{\tau=0} = \frac{4\pi R_c^2}{3} \lim_{\tau \to 0} \left( \tau^{-2p} \pdv{J}{\tau} \right) .
\end{equation}

Finally, the boundary condition \eqref{eq:J_BC_at_x0} can be simplified by introducing the function \(v(\tau,x) = J - \frac{1}{\eta} \pdv{J}{x}\).
Due to linearity (at \(x \ne 0\)), the PDE for \(v\) is the same as for \(J\).
The full PDE problem is then
\begin{equation}
\begin{split}
\label{eq:PDE_and_BC_for_v}
    & \half \pdv[2]{v}{x} + \frac{\tau^{2p}}{3} \pdv{}{\tau} \left( \tau^{2p} \pdv{v}{\tau} \right) = 0 , \\
    & v(\tau\to\infty,x) = 0, \ v(\tau=0,x) = 0, \ v(\tau,x=0) = B .
\end{split}
\end{equation}
Once \(v\) is solved, \(J\) can be calculated by
\begin{equation}
\label{eq:from_v_to_J}
    J(\tau,x) = \eta \int_x^\infty e^{-\eta(y-x)} v(\tau,y) \dd{y} .
\end{equation}

\subsection{Isothermal planar geometry: \(p=0\)}

The simplest case is planar geometry \(p=0\) and constant temperature \(B = \text{const}\).
The system~\eqref{eq:PDE_and_BC_for_v} becomes
\begin{equation}
\begin{split}
    & \half \pdv[2]{v}{x} + \third \pdv[2]{v}{\tau} = 0 , \\
    & v(\tau\to\infty,x) = 0, \ v(\tau=0,x) = 0, \ v(\tau,x=0) = B ,
\end{split}
\end{equation}
whose solution is
\begin{equation}
    v(\tau,x) = \frac{2}{\pi} B \arctan \left( \sqrt\frac{3}{2} \frac{\tau}{x} \right)
\end{equation}
The flux for the \(v\)-problem can be calculated from equation~\eqref{eq:luminosity_from_u}
\begin{equation}
    F_v(x) = \frac{L_v(x)}{4\pi R_c^2} = \frac{2}{3\pi} B \left[ \pdv{}{\tau} \arctan\left(\sqrt\frac{3}{2} \frac{\tau}{x}\right) \right]_{\tau=0}
    = \sqrt\frac{2}{3} \frac{B}{\pi x}.
\end{equation}
The actual flux requires transforming from \(v\) to \(J\) (using equation~\eqref{eq:from_v_to_J}):
\begin{equation}
    F(x) = \sqrt\frac{2}{3} \frac{\eta B}{\pi} \int_x^\infty \frac{e^{-\eta(y-x)}}{y} \dd{y}
    = \sqrt\frac{2}{3} \frac{\eta B}{\pi} e^{\eta x} E_1(\eta x) .
\end{equation}
This agrees with equation~\eqref{eq:simple_scattered_exp_integral} from section~\ref{subsec:flat_isothermal} in the \(x \gg 1\) limit.
It is worth mentioning that for \(x \ll 1\) the exponential integral \(E_1(\eta x)\) diverges logarithmically;
this is a testament to the weakness of the diffusion approximation in describing a single scattering.
The divergence comes from nonphysical scattering states between the line \(\delta\)-function and a single scattering.

\subsection{Isothermal spherical geometry with power-law density: \(p>0\)}

For a constant \(B\), the system~\eqref{eq:PDE_and_BC_for_v} can be solved by noticing that neither the boundary conditions nor the PDE has a length scale in \((\tau,x)\) space, so the solution must be self-similar.
The self-similar variable is \(\xi = \frac{\tau}{x}\) (representing the angle in \((\tau,x)\) space).
Plugging in the ansatz \(v(\tau,x) = B V(\xi)\) gives the ODE
\begin{equation}
    \left( \frac{\xi^2}{2} + \third \right) V''
    + \left( \xi - \frac{2p}{3\xi} \right) V' = 0 .
\end{equation}
The solution for \(V'\) is
\begin{equation}
    V'(\xi) = C \frac{\xi^{2p}}{(3\xi^2+2)^{p+1}} .
\end{equation}
Together with the boundary conditions \(V(0)=0\) and \(V(\infty)=1\), the solution is
\begin{equation}
    V(\xi) = \frac{2}{\sqrt\pi} \frac{\Gamma(p+1)}{\Gamma(p+\half)} \int_0^{\sqrt\frac{3}{2}\xi} \frac{t^{2p} \dd{t}}{(t^2+1)^{p+1}}
    \xrightarrow{\xi \ll 1}
    \frac{2}{\sqrt\pi} \frac{\Gamma(p+1)}{\Gamma(p+\half)} \parfrac{3}{2}^{p+\half} \frac{\xi^{2p+1}}{2p+1}
\end{equation}
The \(\xi \ll 1\) limit is relevant for calculating the outgoing luminosity.
Using equation~\eqref{eq:luminosity_from_u} for the \(v\)-luminosity:
\begin{equation}
    L_v(x) = \frac{4 \pi R_c^2 B}{3} \lim_{\tau\to0} \left[ \tau^{-2p} \pdv{V}{\tau} \right]
    =  4 \pi R_c^2 B \cdot C_p x^{-(2p+1)}
\end{equation}
where \(C_p = \frac{1}{\sqrt\pi} \parfrac{3}{2}^{p-\half} \frac{\Gamma(p+1)}{\Gamma(p+\half)}\).
Performing the \(v \to J\) transformation (equation~\eqref{eq:from_v_to_J}), we get the luminosity spectrum:
\begin{equation}
\label{eq:diffusion_p_solution}
\begin{split}
    L(x) &= 4 \pi R_c^2 B \cdot C_p
    \eta \int_x^\infty \frac{e^{-\eta(y-x)} \dd{y}}{y^{2p+1}}
    = 4 \pi R_c^2 B \cdot C_p \eta x^{-2p} e^{\eta x} E_{2p+1}(\eta x)
    \\
    &= 4 \pi R_c^2 B \cdot C_p \begin{cases}
        \frac{\eta}{2p} x^{-2p} & \eta x \ll 1 \\
        x^{-(2p+1)} & \eta x \gg 1
    \end{cases},
\end{split}
\end{equation}
where \(E_n(z) = \int_1^\infty \frac{e^{-zt}}{t^n} \dd{t}\) is the exponential integral function.
This is in agreement with the rough estimate in section~\ref{subsec:schematic_line_nonflat} (equation~\eqref{eq:rough_line_spectrum_changing_rho}).
Note that this solution diverges for \(x\to0\);
that is due to the diffusion limit treatment of the spectral random walk.

\begin{figure*}
    \centering
    \includegraphics[width=\linewidth]{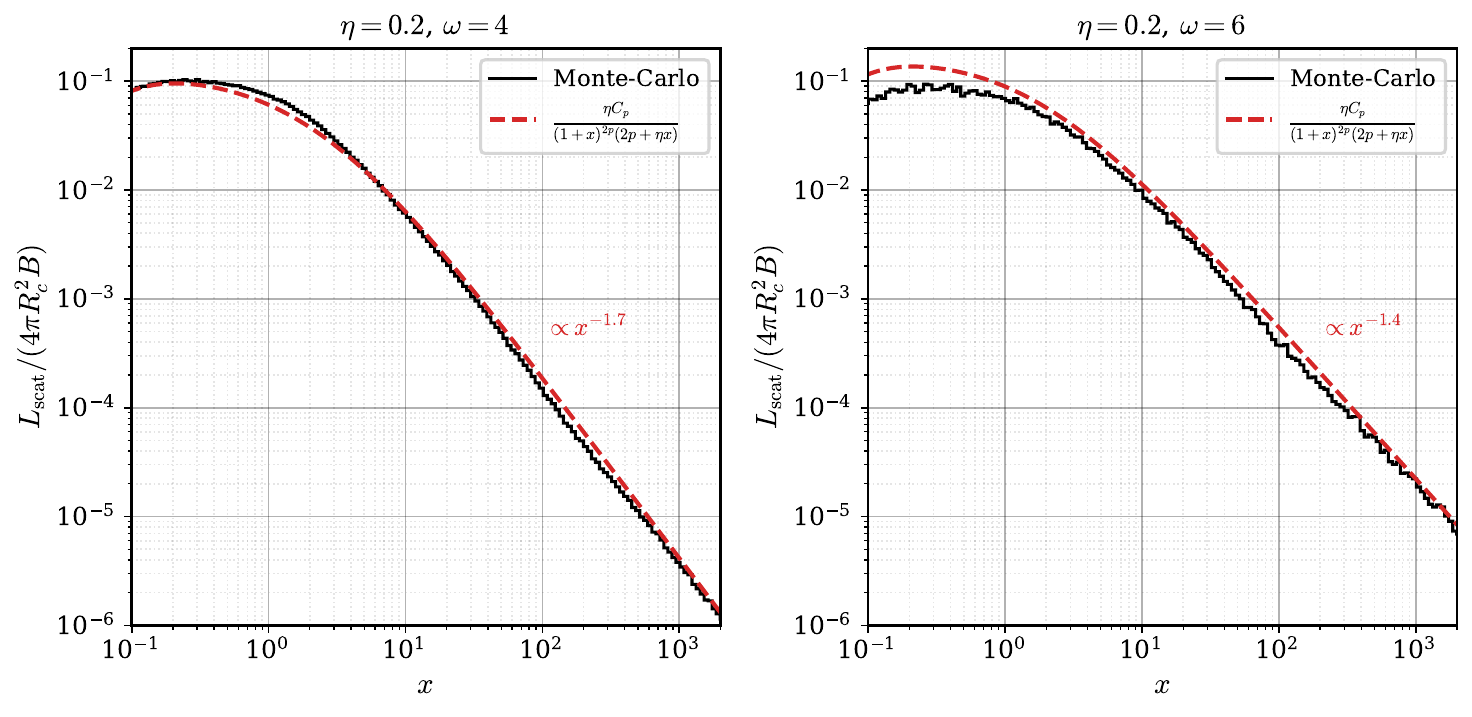}
    \caption{
    Comparison of the scattered line shape in Monte Carlo simulations with \(\eta=0.2\) and \(\omega=4,6\) vs. the simple analytical formula~\eqref{eq:simplest_sph_formula}.
    In the simulations, only photons that have been scattered at least once are counted.
    The analytical formula~\eqref{eq:simplest_sph_formula} is also multiplied by the factor \(\frac{\kps}{\kps + \kappa_\text{line}(x)}\) to account for the nonzero width of the absorption line;
    this is the reason for the dip that starts at \(x \lesssim \eta\).
    }
    \label{fig:sph_sims_vs_analytical}
\end{figure*}

We can construct a simple rational approximation that satisfies both the \(\eta x \gg 1\) asymptotic and the \(\eta x \ll 1\) asymptotic for \(x \gg 1\), and also cancels the non-physical divergence at \(x \ll 1\)
\begin{equation}
\label{eq:simplest_sph_formula}
    L(x) \approx 4 \pi R_c^2 B \frac{\eta C_p}{(1 + x)^{2p} (2p + \eta x)} .
\end{equation}
This fits Monte Carlo simulations reasonably well, as can be seen in figure~\ref{fig:sph_sims_vs_analytical}.

\subsection{Continuum emission}

For the continuum emission, we use equation~\eqref{eq:general_diffusion} with \(\pdv{J}{x}=0\).
\begin{equation}
     \frac{\tau^{2p}}{3} \dv{}{\tau} \left( \tau^{-2p} \dv{J}{\tau} \right) = \beta (J - B) .
\end{equation}
As discussed in section~\ref{subsec:schematic_line_nonflat}, the absorption opacity is generally proportional to density.
If density varies by \(\rho \propto r^{-\omega}\), then \(\beta \propto \tau^{p+1}\) (where, as before, \(p=\frac{1}{\omega-1}\)).
We can be more general and allow \( \beta = \beta_c \tau^\lambda\).
To avoid a divergence of the emission source \(\beta r^2 \propto \tau^{\lambda-2p}\) at \(\tau=0\), we must require \(\lambda > 2p-1\).
The ODE to solve is
\begin{equation}
    \dv{}{\tau} \left( \tau^{-2p} \dv{J}{\tau} \right) = 3\beta_c \tau^{\lambda-2p} (J - B),
\end{equation}
with the boundary conditions of \(J(\infty)=B\) and \(J(0)=0\).
The luminosity is
\begin{equation}
\begin{split}
    L &= \frac{4\pi R_c^2}{3} \lim_{\tau\to0} \tau^{-2p} \dv{J}{\tau}
    = -\frac{4\pi R_c^2 B}{3} (3\beta_c)^{\frac{2p+1}{\lambda+2}} \lim_{s\to0} s^{-2p} \dv{u}{s}
\end{split}
\end{equation}
where we defined the variable \(s = (3\beta_c)^{\frac{1}{\lambda+2}} \tau\) and the function \(u(s) = 1 - \frac{J}{B}\).
\(u(s)\) is the solution of a boundary-value problem that depends only on \(p\) and \(\lambda\):
\begin{equation}
    \dv{}{s} \left( s^{-2p} \dv{u}{s} \right) = s^{\lambda-2p} u, \quad
    u(0)=1, \quad u(\infty)=0.
\end{equation}
This can be solved in terms of the modified Bessel functions, which gives the final result
\begin{equation}
    L = \frac{4\pi R_c^2 B}{3} (3\beta_c)^\nu (\lambda+2)^{1-2\nu} \frac{\Gamma(1-\nu)}{\Gamma(\nu)}, \quad
    \nu = \frac{2p+1}{\lambda+2}.
\end{equation}
If \(p\) and \(\lambda\) came from a power-law density \(\rho \propto r^{-\omega}\), then \(\nu = \frac{2p+1}{p+3} = \frac{\omega+1}{3\omega-2}\), in agreement with the rough estimate of section~\ref{subsec:schematic_line_nonflat}.

\subsection{Line center emission}
\label{subsec:line_center_radially_extended}

The diffusion model only describes the scattered component of the line emission, but the core of the emission line is dominated by non-scattered photons.
Defining
\begin{equation}
    \gamma_x = \frac{\beta_x}{1-\beta_x} = \frac{\kpax}{\kpsx},
\end{equation}
the luminosity of escaping non-scattered radiation is
\begin{equation}
    L_\text{non-scat}(x) = \int_0^\infty 4\pi r^2 B \gamma_x \dd{\tau} \int_{-1}^1 \half e^{-(1+\gamma_x)\tau_\text{esc}(\tau,\mu)} \dd{\mu},
\end{equation}
where \(\tau\) is the scattering optical depth, \(\mu = \cos\theta\) where \(\theta\) is the angle between the radial direction and the photon direction at its creation, and \(\tau_\text{esc}(\tau,\mu)\) is the scattering optical depth over a straight line parallel to that direction from the creation position to infinity.
For a power-law density relation between radius and optical depth \(r = R_c \tau^{-p}\),
\begin{equation}
    \tau_\text{esc}(\tau,\mu) = \frac{\tau}{p} \int_0^\infty \left( 1 + 2 \mu s + s^2 \right)^{-\frac{p+1}{2p}} \dd{s} = \tau g(\mu) .
\end{equation}
We get
\begin{equation}
    L_\text{non-scat}(x) = 2 \pi R_c^2 B \gamma_x (1+\gamma_x)^{2p-1} \Gamma(1-2p) \int_{-1}^1 g(\mu)^{2p-1} \dd{\mu}.
\end{equation}
The inner integral can be solved exactly, yielding
\begin{equation}
\label{eq:non_scattered_general_p}
    L_\text{non-scat}(x) = \pi R_c^2 B \gamma_x (1+\gamma_x)^{2p-1} \Gamma(1-2p) \left[\frac{\sqrt\pi \Gamma\parfrac{1}{2p}}{p \Gamma\parfrac{p+1}{2p}}\right]^{2p} .
\end{equation}
In the flat limit \(p\to0\), this is \(\pi R_c^2 B \frac{\gamma_x}{1+\gamma_x} = \pi R_c^2 B \beta_x\) --- the same as in equation~\eqref{eq:non_scattered_component}.
For \(x \ll \eta\), where \(\gamma_x \gg 1\), the non-scattered emission goes as \(\sim \gamma_x^{2p}\), due to the \(r^2 \sim \tau^{-2p}\) factor for a photon emitted from \(\tau \sim \frac{1}{\gamma_x}\).

\section{Monte Carlo simulations}
\label{sec:monte_carlo}

To validate the analytical analysis of the scattered line profile, we implemented a simple Monte Carlo simulation for the scattering photosphere.
The simulation generates photons with a uniform \(\tau\) probability density for the flat case, and proportional to \(r^2\) in the radially extended case.
The frequency probability density is proportional to the line opacity, and the initial direction is isotropic.
For simplicity, we use a truncated Lorentzian tail \(\frac{\kappa_\text{line}(x)}{\kps} = \min\left[K, \parfrac{\eta}{\pi x}^2\right]\).
The actual value of \(\eta\) is, by definition
\begin{equation}
    \eta_\text{true} = \int \frac{\kappa_\text{line}(x)}{\kps + \kappa_\text{line}(x)} \dd{x}
    = \eta \left[ 1 + \frac{2}{\pi} \frac{K^\half}{K+1} - \frac{2}{\pi} \tan^{-1} \left( K^{-\half} \right) \right]
    \approx \eta \left[ 1 - \frac{4}{3\pi} K^{-\frac{3}{2}} \right] .
\end{equation}
With the usually chosen value of \(K=50\), \(\eta\) is consistent up to \(\sim0.1\%\).

Every generated photon moves in a straight line until it is either electron-scattered, absorbed, or reaches a physical boundary (either the inner boundary or the surface).
In every electron scattering, the velocity of the electron is drawn from a Maxwellian distribution, and the scattering angle \(\theta\) is drawn from a dipole distribution \(\propto 1 + \cos^2\theta\).
The photon's frequency shift due to scattering comes from the Doppler effect
\begin{equation}
    c\frac{\Delta \nu}{\nu_0} = v_x (\cos\theta - 1) + v_y \sin\theta ,
\end{equation}
where \(v_x\) is the electron's velocity component in the original photon direction, and \(v_y\) is the component in the perpendicular direction inside the scattering plane.

To speed up the calculation for photons that are deep in the photosphere and go through many scatterings, a ``super-step'' is implemented, where \(N\) steps are combined into one using the diffusion solution.
\(N\) is determined as the minimum of:
\begin{itemize}
    \item \(0.2 \tau\) --- to limit the chance of reaching the surface.
    \item \(0.2 |x|\) --- to limit the chance of reaching the line core.
    \item \(\left(0.01 (\omega-1) \tau\right)^2\) --- to limit the change in radius (when applicable).
\end{itemize}
A super-step is applied only if the number of steps calculated this way is greater than 20;
otherwise, a single step is taken.


\bibliography{references}{}
\bibliographystyle{aasjournalv7}

\end{document}